\documentclass[prl,aps,onecolumn,showpacs]{revtex4}
 \newcommand{\mytitle}[1]{
 \twocolumn[\hsize\textwidth\columnwidth\hsize
 \csname@twocolumnfalse\endcsname #1 \vspace{1mm}]}
 \newcommand{\beq}{\begin{equation}}
 \newcommand{\eeq}{\end{equation}}
 \newcommand{\bea}{\begin{eqnarray}}
 \newcommand{\eea}{\end{eqnarray}}

\usepackage{graphics}
\usepackage{graphicx}
\usepackage{amssymb}
\usepackage{bm}

\begin{document}

\title{Kondo effect in a side-coupled double-quantum-dot system embedded in a mesoscopic ring}
\author{I-Ling Tsai$^{1}$ and Chung-Hou Chung$^{1,2}$}
\affiliation{ $^{1}$Electrophysics Department, National Chiao-Tung
University, HsinChu, Taiwan, R.O.C. 300\\
$^{2}$ Department of Physics, Yale University, New Haven, Connecticut, U.S.A., 06520
}
\date{\today}
\begin{abstract}
We study the finite size effect of the Kondo screening cloud in a
double-quantum-dot setup via a large-N slave-boson mean-field theory. In
this setup, one of the dots is embedded in a close metallic ring
with a finite size $L$ and the other dot is side-coupled to the
embedded dot via an anti-ferromagnetic spin-spin exchange
coupling with the strength $K$. The antiferromagnetic coupling
favors the local spin-singlet and suppresses the Kondo screening.
The effective Kondo temperature $T_k$ (proptotional to 
the inverse of the Kondo
screening cloud size) shows the Kosterlitz--Thouless (KT) scaling
at finite sizes, indicating the quantum transition of the 
KT type between the
Kondo screened phase for $K \le K_c$ and the local spin-singlet
phase for $K\ge K_c$ in the thermodynamic limit with $K_c$ being 
the critical value. 
The mean-field phase diagram as a function of $1/L$ and $K$ shows 
a crossover between Kondo and local spin-singlet ground states 
for $K<K_c$ ($L=4n, 4n+1, 4n+3$) and for $K>K_c$ ($L=4n+2$).
To look into the crossover region more closely, the 
local density of states on the quantum dot 
 and the persistent
current at finite sizes with different values of 
$K$ are also calculated. 
\end{abstract}


\maketitle

\subsection{\large \bf I. Introduction}

Kondo effect\cite{Hewson}, the screening of magnetic impurities by 
the conduction electron reservoir, has been intensively 
studied over the last half-century.  
  Recently, there has been revival interest in Kondo effect in 
semiconductor  quantum dot devices due to the progress in 
fabracating nanostructures\cite{goldharber}. When there are odd number of electrons 
on the dot, Kondo effect overcomes the Coulomb Blockade, 
leading to a narrow pronounced resonance peak in local impurity 
density of states and enhanced conductance through the quantum 
dot\cite{glazman}\cite{goldharber}. The Kondo effect 
is characterized by the single energy scale 
, the Kondo temperature $T_k\approx D e^{-1/J}$\cite{Hewson}, 
below which the Kondo
 screening develops. Here, $D$, $J$ in $T_k$ 
are the conduction electron bandwidth 
and the dimensionless Kondo coupling, respectively. 
In a finite-sized mesoscopic device, $T_k$ sets the length scale 
$\xi_k^0 \approx \hbar v_F/T_k$ associated with the size of the Kondo 
screening cloud where the cloud of electrons with the size of order 
of $\xi_k^0$ surrounding the magnetic impurity form spin-singlet state 
with it\cite{ring1}. Here, $v_F$ is the Fermi velocity. 
For typical values of $T_k$, $\xi_k^0$ is around $0.1~ 1\mu m$, 
which is comparable to the typical size of the quantum dot device, 
leading to the finite size effect of the Kondo screening cloud.
This effect has been investigated in a quantum dot embedded in a 
mesoscopic ring threaded by a magnetic field\cite{ring1}\cite{ring2}\cite{ring3}. 
The experimentally measurable 
persistent current induced by the magnetic flux has been shown to be 
sensitive to the ratio of the size of the ring $L$ and $\xi_k^0$
\cite{ring1}\cite{ring2}\cite{ring3}. 
As $L \ll \xi_{K}^0$ the Kondo screening cloud does not develop completely, 
giving rise to the suppression of Kondo effect and a reduction of the persistent current 
even if the temperature is lower than $T_k$; while as for $L\gg \xi_k^0$ 
the Kondo screening cloud is formed, leading to enhanced persistent current. 
By measuring the transport 
properties (such as: persistent current) 
as a function of the system size, we gain insights on how 
the Kondo screening cloud is formed as the system size is increased. 
This idea offers another route to realize Wilson's  
Numerical Renormalization Group (NRG)\cite{NRG} idea on the Kondo problem. 

Very recently, study of the Kondo effect has been extended to 
the coupled double-quantum-dot setups via antiferromagnetic 
RKKY interactions\cite{Marcus2QD}\cite{chung2QD}\cite{rkkyPascal}
\cite{cornaglia}.  
The Kondo effect in such double-dot systems 
competes with the RKKY interactions, 
giving rise to quantum phase transition in the context of 
the well-known two-impurity 
Kondo problem between the Kondo and 
local spin-singlet ground states\cite{2impkondo}. 
Close to the quantum critical point physical 
observables at finite temperatures exhibit 
non-Fermi liquid behaviors. 
The crossover behaviors between these two phases 
can be accessed by changing temperatures. 
However, at zero temperature but at finite-sizes, 
the size of the Kondo screening cloud provides us with 
an alternative route to the crossover behaviors in the two-impurity 
Kondo problem\cite{pascal}.

In this paper, we study the Kondo effect
 in a side-coupled double quantum dot system 
embedded in a finite-sized mesoscopic ring. 
A similar 
side-coupled double-quantum-dot system has been studied  
where one of the dot 
is coupled to Fermi-liquid leads of conduction electrons 
with continuous spectrum\cite{side-coupled}. 
Unlike the two-impurity Kondo system, in the side-coupled 
double-dot system the Kondo phase is fragile and 
unstable towards the local spin-singlet state for 
any infinitesmall antiferromagnetic spin-exchange coupling, 
and the transition is of the 
Kosterliz-Thouless type. Note that a related setup consisting 
of double quantum dots in parallel connected to a mesoscopic 
ring shows a quantum phase transition of the pseudogap 
Anderson model when the magnetic flux is tuned\cite{dias}. 
Here, we are interested in not only 
the nature of the 
Kondo-to-spin-singlet quantum phase transition in our proposed setup 
but also the crossover phenomena via a systematic 
study of the system at finite sizes. The inverse of the system size 
$1/L$ plays a similar role as temperature $T$. The finite 
temperature crossover behaviors between the above two quantum ground states 
can therefore be accessed effectively via changing the size of the ring.

We focus on the following three quantities to investigate this issue: 
the effective Kondo temperature $T_k$ (inversely proportional to the 
size of the Kondo screening cloud $\xi_k$ 
in the presence of the antiferromagnetic spin exchange coupling), 
the local density of states 
(DOS) $\rho_{QD}(\omega)$ on the quantum dot, and the persistent current (PC) $I$. The plan of the paper is as follows. In Section II., we introduce the model and present a large-N mean-field treatment of the model. In Section III., we present our results on the effective Kondo temperature, the local DOS on the 
dot and the persistent current.
We also give detail explanations of our results. The conclusions are given in Section IV..


\vspace{0.15cm}
\begin{figure}[h]
\begin{center}
\includegraphics[width=7.5cm]{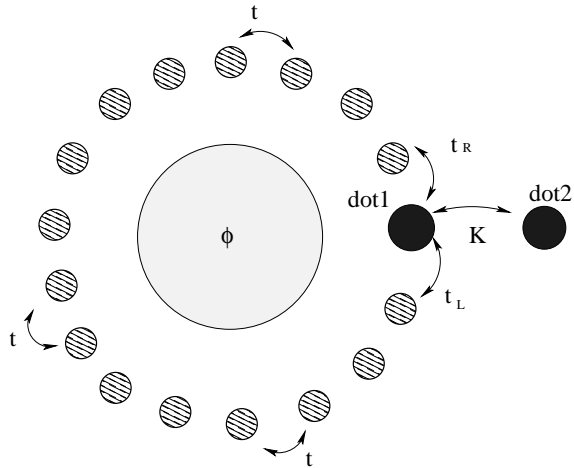}
\end{center}
\vspace{0.05cm}
\label{setup}
\caption{Sketch of the model. $t_L / t_R $ is the hopping coefficient 
between dot $1$ and its left/right conducting island on the ring, 
$\phi$ is the magnetic flux (in unit of $2\pi/\Phi_0$ with $\Phi_0=hc/e$) through the ring, $K$ is antiferromagnetic coupling between dot $1$ and dot $2$.}
\end{figure}

\subsection{\large \bf II. Model}

Our model describes a double-quantum-dot system embedded in electronic 
reservoir in a finite-sized ring (see Fig.\ref{setup}). 
In this setup, only one of the dot (dot $1$) is coupled to  
the ring via hoping; while the other dot (dot $2$) is coupled 
only to dot $1$ via antiferromagnetic spin-exchange 
interaction with a coupling strength $K$ and is 
decoupled from the reservoir of the ring. The ring serves as electronic 
reservoir consists of $L-1$ conducting islands. We 
consider here the half-filled case ($N_e =L+1$ where $N_e$ is the total 
number of electrons on the ring and on the two dots). 
The key point in this setup with a 
closed geometry 
is that the Kondo screening cloud is trapped in the ring and can not 
escape into the external leads\cite{ring1}. Nevertheless, one can measure 
the transmission probability through the quantum dot $1$ embedded in the ring by measuring the persistent current\cite{ring1}\cite{Buttiker} (see details in Sec. III.). 
Here we assume a direct antiferromagnetic spin exchange coupling between 
two quantum dots which competes with the Kondo effect. 
Note that though it has been known experimentally that 
the antiferromagnetic spin-spin coupling (RKKY interaction) 
between a double-quantum-dot system can be 
induced naturally by a conduction electron island in the middle of the two dots\cite{Marcus2QD}, the direct antiferromagnetic spin exchange 
coupling may in principle be generated via the second-order 
hoping process directly between the two dots 
if the two quantum dots are close 
enough to each other.  
Since in both cases we expect 
suppression of the Kondo effect due to local spin-spin exchange 
coupling, to simplify our calculations we assume a direct antiferromagnetic 
spin-exchange coupling between the two dots.

The Hamiltonian of our model is given by:

\begin{eqnarray}
&{H}& = H_{A}+ KS_{1} \cdot S_{2}
\end{eqnarray}

where $H_{A}$ represents a quantum dot embedded in a ring by the 
Anderson impurity model. 

\begin{eqnarray}
{H}_{A}  &=&   -  t\sum_{j=1}^{L-1} \sum_{\sigma}[c_{j}^{\dagger}
c_{j+1}+ H.C] + U\sum_{i=1,2}n_{d_i}^{\uparrow}n_{d_i}^{\downarrow} 
+ \sum_{\sigma,i=1,2} \epsilon_{d}d_{i\sigma}^{\dagger}d_{i\sigma}\\ \nonumber
& - & \sum_{\sigma} [t_{L}d_{1\sigma}^{\dagger}c_{1\sigma} + t_{R}\exp
(i\phi) c_{L-1\sigma}^{\dagger}d_{1\sigma}+ H.c.]
\end{eqnarray}
where $c_{j}$, $d_1$, $d_2$ represent the annihilation operators 
of electron on site $j$ of the ring, dot $1$ and dot $2$, respectively,
 the phase factor $\phi$ is defined by 
$\phi=2\pi \Phi/\Phi_0$ with $\Phi$ being 
the magnetic flux going through the ring and $\Phi_0=hc/e$ being 
the flux quantum, 
$t$ is the electron hoping within the tight-binding ring, and  
$t_{L/R}$ are the hopings between dot $1$ and the two neighboring 
ring electrons ($c_{L-1}$/$c_{1}$). 
Here,  $S_{1}$ and $S_{2}$ represent spin operators of dot $1$ 
and dot $2$, respectively with $S_{i}=\sum_{\sigma
\sigma'}d_{i\sigma}^{\dagger}\sigma_{\sigma\sigma'}d_{i\sigma'}$. 
$L$ is the total number of electrons in the system
excluding dot $2$.
The well-known Kondo limit is reached for $U\gg t$.
Here, we consider a simple limit in the Kondo regime, 
$U \rightarrow \infty$ where the slave-boson mean-field (SBMF) 
approach\cite{ring2}\cite{ring3} is applicable to simplify the 
quartic $U$ term.

Further progress can be made by decoupling the quartic 
spin-spin interaction $S_1\cdot S_2$ 
into qudratic one via the Hubbard-Stratonovish transformation 
in the framework of the large-N $SU(N)$ mean-field theory where 
the symmetry of the Hamiltonian is generalized from the $SU(2)$ with 
spin degeneracy being two to 
$SU(N)$ with spin degeneracy being $N\rightarrow\infty$\cite{Hewson}.   
In the SU(N) generalization of the 
slave-boson representation, we have 
$d_{i \alpha}$ = $ f_{i\alpha}b_{i}^{\dagger}$, 
$\alpha$ is the flavor of the spin: $\alpha = 1,2,...N$, $i=1,2$ is the 
indix for the two quantum dots.  
The local constraints to enforce the single occupancy on dot $1$ 
and $2$ are given by
\begin{eqnarray}
\sum_{\alpha} f_{i}^{\dagger \alpha}f_{i \alpha} + b_{i}^{\dagger}b_{i} = \frac{N}{2}
\end{eqnarray}

The mean-field Hamiltonian for $H_A$ is therefore given by,
\begin{eqnarray}
N H_{A, MF}  &=&   - 
t\sum_{j=1}^{L-1}\sum_{\alpha=1}^{N}[c_{j\alpha}^{\dagger}
c_{j+1\alpha} + H.c.] + \sum_{\alpha=1}^{N} \widetilde{\epsilon_{d}}
f_{1 \alpha}^{\dagger}f_{1 \alpha}  + \sum_{\alpha=1}^{N} \epsilon_{d}^{\prime}
f_{2 \alpha}^{\dagger}f_{2 \alpha}\\ \nonumber
& - & \sum_{\alpha=1}^{N} [\widetilde{t_{L}}
f_{1\alpha}^{\dagger}c_{1\alpha}
+ \widetilde{t_{R}}\exp(i\phi)
 c_{L-1\alpha}^{\dagger}f_{1 \alpha}+ H.c] + 
 \lambda(b_0^2-1) + \bar{\lambda} (\bar{b}^2-1)
\end{eqnarray}

where $\widetilde{t}_{L/R} = b_0 t_{L/R}$ with $b_0$ being the expectation value 
of the $b_1$ boson on dot $1$, $b_0=<b_1>$, $\bar{b}=<b_2>$, 
$\lambda$ ($\bar{\lambda}$) is the Lagrange multiplier which enforces the local contraint on dot $1$ ($2$), $\tilde{\epsilon_d}=\epsilon_d-\lambda$, $\epsilon_d^{\prime}=\epsilon_d-\bar{\lambda}$. 
The quartic antiferromagnetic spin-spin interaction 
can be decoupled via the mean-field variable $\chi$
\cite{chung-largeN}:
\begin{eqnarray} 
\chi = \frac{b_0 K}{N} <d_{1 }^{ \dagger \alpha }d_{2, \alpha}>
\end{eqnarray}. Therefore,
\begin{eqnarray}
KS_1\cdot S_2 &=& -\sum_{\alpha=1}^{N}\frac{\chi}{N}  
f_1^{\dagger \alpha} f_{2 \alpha}-H.c.+ \frac{\chi^2}{K}
\end{eqnarray}
The full mean-field Hamiltonian is given by
\begin{eqnarray}
H_{MF} = H_{A, MF} -\sum_{\alpha=1}^{N}\frac{\chi}{N}  
f_1^{\dagger \alpha} f_{2 \alpha}-H.c.+ \frac{\chi^2}{K}.
\end{eqnarray}

The mean-field energy $E_{MF}$ can be obtained by  
diagonalizing the above Hamiltonian and can be expressed as
\begin{eqnarray}
E_{MF} = \sum_{m, \sigma}\epsilon_{m, \sigma}(\lambda, b_0) + \frac{\chi^2}{K} 
+ \lambda(b_0^2-1) + \bar{\lambda} (\bar{b}^2-1),
\end{eqnarray}
where $\epsilon_{m,\sigma}$ are the eigenvalues of the Hamiltonian matrix in 
$H_{MF}$, the summation over $\emph{m}$ includes all occupied levels of
$H_{MF}$. The values of the mean-field variables 
$\lambda$, $b_0$ and $\chi$ are determined by minimizing $E_{MF}$ 
with respect to $b_0$ and $\chi$: 
\begin{equation} 
\frac{\partial E_{MF}}{\partial b_0} = \frac{\partial E_{MF}}{\partial \bar{b}} = \frac{\partial E_{MF}}{\partial \chi} =0,
\end{equation}
subject to the constraint $\frac{\partial E_{MF}}{\partial \lambda} = \frac{\partial E_{MF}}{\partial \bar{\lambda}}=0$.
The ground state energy $E_{gs}$ 
corresponds to the global minimum of 
$E_{MF}(\lambda,\bar{\lambda}, b_0, \bar{b}, \chi)$ which satisfies the 
mean-field equations. 
Note that the advantages of taking the large-$N$ mean-field approach are: 
(i) in the large-$N$ limit the solutions from the mean-field equations 
are exact though the physical system corresponds to $N=2$, 
and (ii) at finite $N$, a systematic $1/N$ correction to the mean-field 
results is possible though it is beyond the scope of this paper.   
The mean-field phase diagram can be mapped out via the solutions of the 
above mean-field equations. 

\begin{figure}[h]
\begin{center}
\includegraphics[width=11cm]{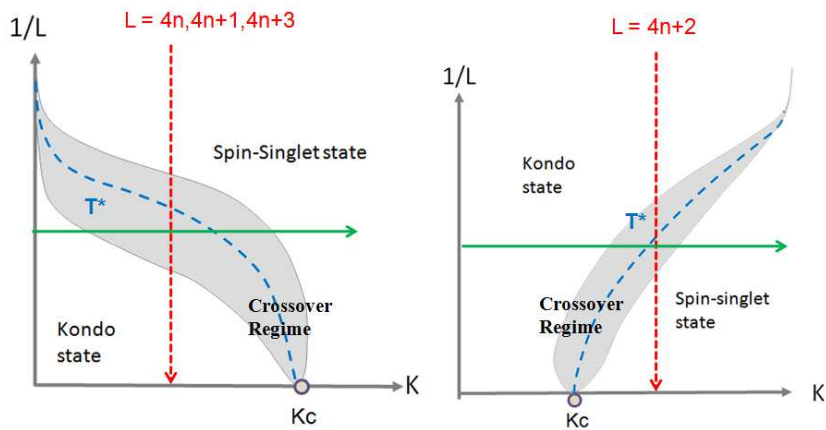}
\end{center}
\caption{Schematic mean-field 
phase diagram of the model for (a) $L=4n, 4n+1, 4n+3$ (case (I)) 
and (b) $L=4n+2$ (case (II)). The red line indicates the crossover with $K$ being fixed and the green line indicates the crossover with $L$ being 
fixed.}
\label{phase}
\end{figure}

\subsection{\large \bf III. Results}

Three physical observables are 
identified to investigate
 the Kondo screening effect of our setup at finite sizes: 
(i). \emph{Effective Kondo temperature}($T_k$), 
(ii). \emph{Local density of states on dot $1$ (LDOS)} ($\rho_{QD}(\omega)$),  
(iii). \emph{Persistent current(PC)}. Details are shown below.

Before we present our new results, it is useful to 
summarize the behavior of the model at $K=0$ (without 
antiferromagnetic spin-spin coupling), which has been intensively 
studied\cite{ring1}\cite{ring2}\cite{ring3}. 
The Kondo resonance 
of a quantum dot embedded in a mesoscopic ring strongly depends on 
the finite size $L$ (mod $4$), and the magnetic flux $\Phi$ 
threading the ring. 
In particular, both the Kondo temperature $T_k$ and the persistent current 
$I$ follow universal scaling functions  
of $\xi_k^0/L$ at a fixed magnetic flux. In the Kondo regime, all physical 
observables, such as: Kondo temperature $T_k$, persistent current $I$ 
 are enhanced as the size $L$ increases, but with different crossover 
bahaviors in all four cases of $L=4n$, $L=4n+1$, $L=4n+2$, and $L=4n+3$.
 The magnetic flux dependence of persistent current $I$ exhibits a 
symmetry between size $L$ and $L+2$: $I_L(\phi) = I_{L+2}(\phi + \pi)$, 
indicating that adding the magnetic flux of $\Phi_0/2$ (or adding a phase 
$\pi$)  is equivalent to switch the behavior of the PC from a system with 
size $L$ to $L+2$.

With the previous results in mind, we may discuss the general 
properties for $ K > 0$. Due to the antiferromagnetic 
spin-spin couping, we expect in this case 
the competition between the Kondo and the local spin-singlet ground states, 
leading to the quantum phase transition. In fact, quantum phase transitions 
in double-quantum-dot systems with antiferromagnetic RKKY interactions 
have been intensively studied in recent years
\cite{Marcus2QD}\cite{chung2QD} in the framework of two-impurity Kondo problem
\cite{2impkondo} where the quantum dots couple to 
the conduction electron Fermi sea with continuous spectrum. Two types of 
quantum phase transitions have been identified in these systems: 
the phase transition 
with a quantum critical point (QCP) 
and the one of the Kosterlitz-Thouless (KT) type. 
The characteristic behavior near QCP is the observables power-law 
dependence on the coupling strength relative to the critical point; while 
for the KT transition the crossover energy scale exponentially depend 
on the distance to the critical point.   
The former (QCP) type of the quantum phase transition 
is realized in the double-dot systems where each of the dot 
couples to an independent conduction electron 
reservoir\cite{Marcus2QD}\cite{chung2QD}, 
and the critical 
point separating the Kondo from the local spin-singlet phase 
is the well-known two-impurity Kondo fixed point\cite{2impkondo}. The latter 
 (KT) type 
exsits in a side-coupled double-dot system where only one of the dots 
coupled to the electron reservoir\cite{side-coupled}. 
The Kondo resonance becomes more fragile 
in the side-coupled system so that an infinitesmall antiferromagnetic 
coupling is sufficient 
 to suppress the Kondo effect and leads to the spin-singlet ground state.
We expect the similar Kosterlitz-Thouless transition to occur in our 
side-coupled double-dot system embedded in a ring. However, 
the mesoscopic ring in our setup 
consists of a finite number of tight-binding electrons 
(instead of a Fermi sea with continuous spectrum), 
the details of the transition   
might be different from those in  
Ref.\cite{side-coupled} (see below).

\subsubsection{\bf A. Mean-field phase diagram}

After solving the mean-field equations, we summarize our main results in the 
schematic mean-field phase diagram as shown in Fig. \ref{phase}. We find 
indeed the transition between the Kondo and spin-singlet phases is 
of the Kosterlitz-Thouless type (see below). However, we find 
that the critical point $K_c$ 
separating the two phases is not at zero as shown in the similar side-coupled 
double-dot system studied previously in Ref. ~\cite{side-coupled} 
but at a finite value: $K_c> 0$. 

There are three regions in the phase diagram, corresponding to different 
mean-field solutions:\\

1. For small $K$ we find 
$b_0\ne 0$, $\lambda\ne 0$, $\chi=0$. In the thermodynamic limit 
this is the Kondo phase studied in Ref.\cite{ring1}\cite{ring2}\cite{ring3}.\\

2. For large $K$, we find $b_0= 0$, $\lambda\ne 0$.  
The ground state for $L\rightarrow \infty$ 
is the local spin-singlet phase where antiferromagnetic spin-spin 
coupling completely 
suppresses the Kondo effect.\\

3. In the intermediate values of $K$, we find 
$b_0= 0$, $\lambda\ne 0$, $\chi\ne 0$. 
This corresponds to the
 crossover region between Kondo and spin-singlet phases indicated 
in the shaded region in Fig.\ref{phase}. This region is defined 
either by $K_{c1}<K<K_{c2}$ for a fixed size $L$ where 
$K_{c1}$ and $K_{c2}$ are the boundaries between 
the crossover region and the two stable phases or by $L_{c1}<L<L_{c2}$ for a fixed $K$ where 
$L_{c1}$ and $L_{c2}$ are defined in a similar way as $K_{c1}$ and $K_{c2}$.

Note that in all the above three cases, we find $\bar{b}=0$, $\bar{\lambda}=\epsilon_d$. We then systematically investigate the crossover behaviors along 
the following two different 
crossover paths, depending on the finite size $L$ (mod $4$): \\

Case (I) (Fig.\ref{phase} (a)) holds for $L=4n,4n+1,4n+3$ where 
we find the crossover occurs for $K<K_c$.\\ 

Case (II) (Fig.\ref{phase} (b)) 
occurs for $L=4n+2$ where we find the crossover ranges exist mainly for  
$ K> K_c$.\\ 

 In both cases, we find the crossover energy scale $T^*$ follows 
the behavior of the typical Kosterlitz-Thouless transition:
\begin{equation}
T^*= c \widetilde{T_{k}}\exp{[-\pi\widetilde{T_{k}}/(K-K_c)]}
\end{equation}
 where $\widetilde{T_k}=c'T_k$, and both $c$ and $c'$ are non-universal constants.

We investigate the Kondo effect in our setup at finite sizes by either 
changing the size $L$ at a fixed $K$ (or following 
the red (vertical) line in Fig.\ref{phase}) 
or changing $K$ at a fixed size 
$L$ (or following the green (horizontal) line in Fig. \ref{phase}). 
The finite size scaling of $T_k$ indicates that 
in the thermodynamic limit $L\rightarrow \infty$, 
$K_{c1}$ and $K_{c2}$ converge to a single critical value $K_c$:
$K_{c1}=K_{c2}\rightarrow K_c$ (see below).

\subsubsection{\bf B. The effective Kondo temperature}

 In the single Kondo dot system embedded in a 2D electron gas (2DEG), 
it is well-known that the physical 
properties follow universal functions of $(T/T_K^0)$ in Kondo
regime where $T_k^0$ is the Kondo temperature of the single dot 
in the thermodynamic limit\cite{goldharber}. 
In a single quantum dot embedded in a mesoscopic ring with 
a finite size, the effective Kondo temperature 
can be defined as\cite{ring2} 
\begin{eqnarray}
T_k^{1QD}= \epsilon_d-(E_{gs}-E_{gs}^{o})
\label{Tk1QD}
\end{eqnarray}
, where $\epsilon_d$ is the energy of the dot $1$, 
$E_{gs}$ and $E_{gs}^0$ correspond to the energy of the full 
system and of the tight-binding ring with the open boundary condition, 
respectively. The effective Kondo temperature 
$T_k$ is the energy gain due to the coupling between quantum dot and the ring, 
which corresponds to the energy associated with the Kondo coupling. 
In the thermodynamic limit ($L\rightarrow \infty$), 
$T_k$ approaches to $T_k^0$. 
For the relevant parameters in the $K=0$ 
Anderson model of the 
single dot-ring setup: $t=1$, $t_L=t_R=0.4 t$, $\epsilon_d=-0.8 t$, 
we find $T_k^0\approx 0.0189 t$, $\xi_k^0\sim 106$. Note that for simplicity, 
we set $t=1$ throughout the paper as our unit.

In the presence of the antiferromagnetic spin-spin coupling 
the effective Kondo temperature $T_k$ 
at a finite size in the context of our large-N slave-boson mean-field 
approach can be generalized from Eq. \ref{Tk1QD} to:
\begin{eqnarray}
T_k= \epsilon_d-(E_{gs}-E_{gs}^{o})-\frac{\chi^2}{K}
\label{Tk2QD}
\end{eqnarray}
Note that the effective $T_k$ defined in Eq.\ref{Tk1QD} and Eq.\ref{Tk2QD} 
are slightly different from that defined in Ref.\cite{ring2} where 
$T_k$ is measured relative to the hightest occupied level defined 
as $\epsilon_F$, corresponding to the Fermi energy in the thermodynamic 
limit\cite{ring2}\cite{ring3}. 
Though this minor difference in definition 
does not affect the results and the physics, $T_k$ defined here 
offers an 
intuitive understanding of the crossover at finite sizes corresponding 
to case (I) and (II) mentioned above--for $K<K_c$, $T_k$ 
is expected to increase with increasing size $L$ for case (I) 
as the system recovers 
the Kondo resonance in $L\infty$ limit,  
and for $K>K_c$ 
$T_k$ is expected to decrease 
to $0$ as the system approaches to the local spin-singlet 
ground state in the thermodynamic limit 
 (see details below). 
Nevertheless, we have checked that with the definition 
in Ref. \cite{ring2} and \cite{ring3} our results for $T_k$ at finite sizes  
 indeed reproduce those in Ref.\cite{ring2} and \cite{ring3}. 
We have also checked that our results on $T_k$ for $K=0$ 
are consistent 
with the finite-size behaviors in $E_{gs}$ and $E_{gs}^0$ 
via perturbation theory in Ref. \cite{ring1}.     

In the Kondo phase, $T_k$ reduces to the 
Kondo temperature for the single quantum dot embedded in a mesoscopic ring 
($K=0$). When $L \rightarrow  \infty$, $T_k$ in the Kondo phase  
approaches to $T_K^0$.  In the crossover 
between Kondo and the local singlet phases, $T_k$ decreases with increasing 
$K$, and finally $T_k\rightarrow 0$ when the system is in the local singlet 
ground state where the Kondo effect is completely suppressed by the antiferromagnetic spin-spin couping. We analysize below in details 
the crossover between the Kondo and the local spin-singlet ground states 
from the behaviors of the Kondo temperature.

\subsubsection{ \textbf{ 1. Varies K with fixed L }}
\vspace{5mm}
\begin{figure}[h!]
\begin{center}
\includegraphics[width=16cm ]{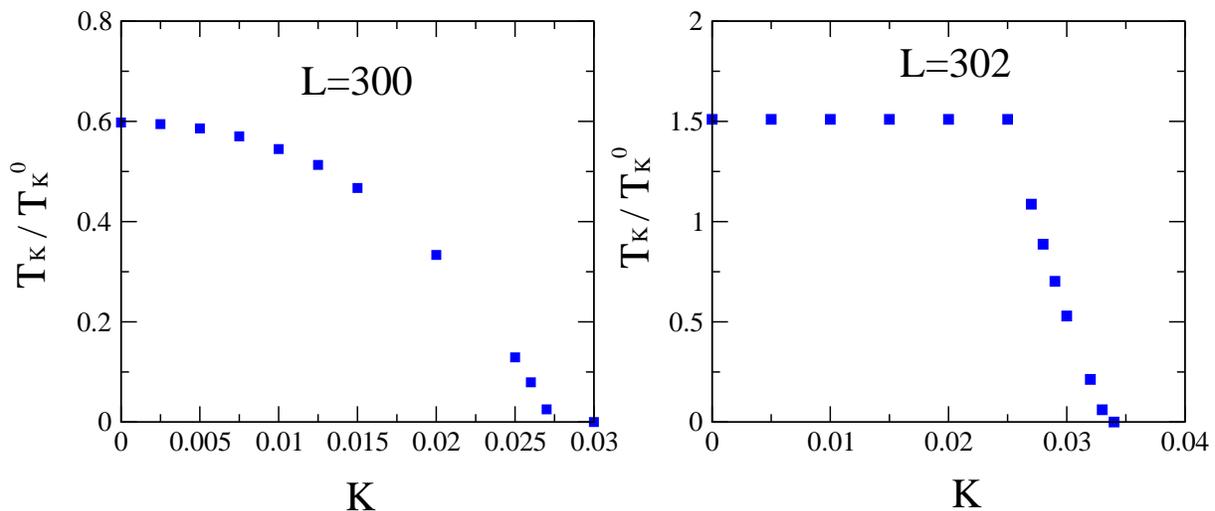}
\end{center}
\caption{$T_k$ as functions of $K$. (a)Case I: $L=4n$ (b)Case II: 
$L=4n+2$.  
Other parameters: $\epsilon_d$ = -0.8t, $t_R$ = $t_L$ = 0.4t, $\phi=0$. 
Here we set $t=1$ as the unit.}
\label{TK-k}
\end{figure}

We first monitor how the Kondo resonance is destroyed by varying 
the antiferromagnetic spin-spin 
coupling strength $K$ at a fixed finite size $L$.  
Fig.(\ref{TK-k}) shows $T_k$ as a function of antiferromagnetic 
coupling strength 
$K$ at a fixed size $L=300$ (or $\xi_k^0/L\sim 0.3$). 
In both case (I) and (II), $T_k$ vanishes as $K\rightarrow K_{c2}$, 
indicating the suppression of Kondo resonance by the antiferromagnetic 
spin-spin interaction. 
 However, there are minor differences between these two cases 
in how fast $T_k$ vanishes 
as $K$ close to $K_{c2}$. 
In case (II), $T_k$ remains a constant over a wider range of $K<K_{c1}$ 
compared to that in case (I) before it decays.  
This suggests that the Kondo effect 
at a finite size 
seems more robust in case (II) ($L=4n+2$) than in case (I) 
($L=4n,4n+1,4n+3$) so that the crossover 
region in case (II) is much narrower than in that in case (I). 
This is consistent with our numerics as $T_k$ for $L=4n+2$ 
is found to have the largest value among $L$ (mod $4$) for a given size.

\vspace{5mm}
\subsubsection{\textbf{ 2. Varies $L$ with fixed $K$}}
\vspace{5mm} 

Next, we present the results on the finite-size 
dependence of the Kondo resonance 
at a fixed antiferromagnetic spin-spin coupling 
strength $K$. In analogous to the Numerical Renormalization Group 
(NRG) method, the ground state is computed and monitored as 
we decrease the energy scale $1/L$ (or equivalently increase 
the system size $L$)  
until we reach the thermodynamic limit $L\rightarrow \infty$ 
(or effective zero temperature). 

\begin{figure}[h!]
\begin{center}
\includegraphics[width=14cm, height=6cm ]{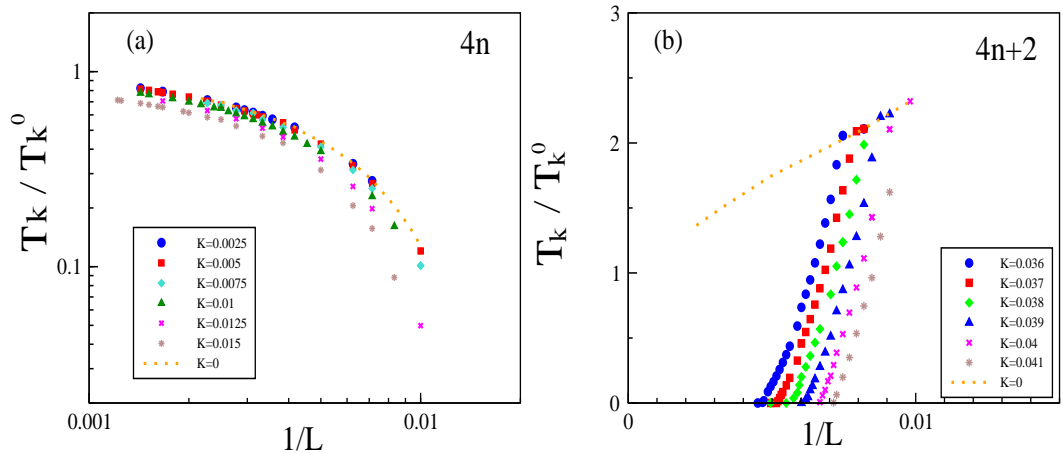}
\end{center}
\caption{$T_K$ as functions of $1/L$: (a) $L=4n$, 
$K$ varies from $0.0025t$ to $0.015t$, distance: 0.0025t. 
(b) $L=4n+2$, $K$  varies from $0.036t$ to
 $0.04t$, distance: $0.001t$. Other parameters: $\epsilon_d$ = -0.8t, $t_R$ = $t_L$ = 0.4t,
 $\phi=0$. Here, we set $t=1$ as the unit.}
\label{TK4n}
\end{figure}

First, we describe the qualitative behaviors of $T_k$ at finite sizes 
in the two cases as mentioned above. 
For case (I) (represented by $L=4n$, see Fig.\ref{TK4n}(a)) 
$T_k$ increases with increasing $L$ as the crossover region 
is for $K<K_c$ where the Kondo resonance is recovered at large 
system size $L\gg\xi_k^0$; while for case (II) ($L=4n+2$, see Fig.\ref{TK4n}(b)) 
since the crossover 
occurs for $K>K_c$, $T_k$ decreases with increasing $L$ and it vanishes 
in the thermodynamic limit.
It is clear from the crossover behavior in Fig. \ref{TK4n} that 
the finite-size effect appears for $L\l\xi_k^0$ where the Kondo screening 
cloud has not yet fully developed for $K<K_c$ and has not yet completely 
destroyed for $K>K_c$; while this effect diminishes as the system approaches to 
the thermodynamic limit or $L\gg \xi_k^0$. 
Note that in case (I) with large $K>K_c$ and case (II) with
small $K<K_c$, the ground state remains at local singlet 
and Kondo state, respectively; therefore, no crossover behaviors 
are found. It is also worthwhile noting that $T_k$ for $L=4n+2$ (case (II)) 
for $K=0$ approaches to $T_k^0$ from above as $L$ approaches to the thermodynamic limit. This suggests that the Kondo effect in this case 
seems more robust 
at finite size than in the thermodynamic limit.  
Therefore, at any finite size $L$, it is necessary to apply a larger antiferromagnetic 
coupling $K$ compared to $K_c$ which is required in $L\rightarrow \infty$ limit to suppress the Kondo effect. This provides 
an explanation why we always find the crossover behavior for $L=4n+2$ 
for $K>K_c$. On the other hand in case (I) ($L=4n,4n+1,4n+3$) for $K=0$, 
$T_k$ approaches to $T_k^0$ from below  
as $L$ increases, which explains why the crossover occurs for $K<K_c$ 
in this case.

\vspace{5mm}
\vspace{5mm}

\begin{figure}[h!]
\begin{center}
\includegraphics[width=14cm, height=6cm ]{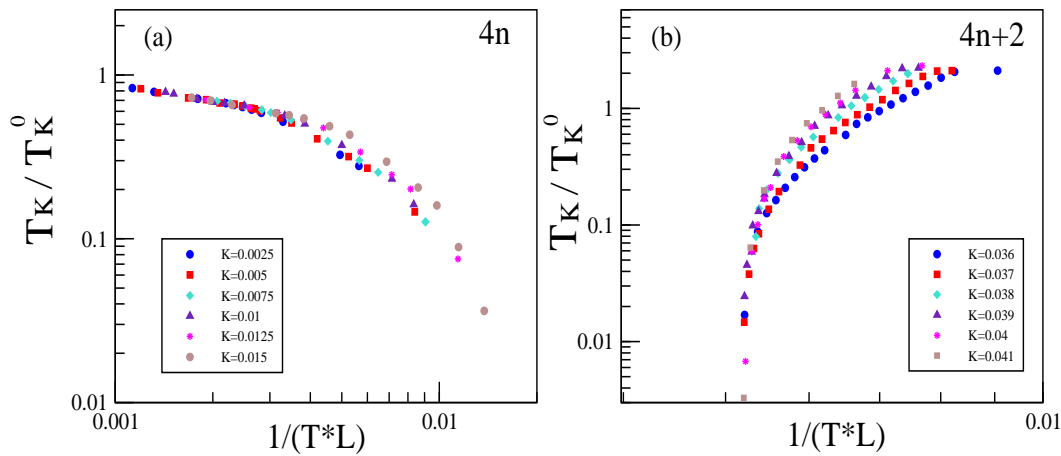}
\end{center}
\caption{$T_K$ as functions of $1/(T^*L)$: (a) $L=4n$, $K$ 
varies from $0.0025t$ to $0.015t$, distance: $0.0025t$. 
(b)  $L=4n+2$, $K$ 
varies from $0.036t$ to $0.04t$, distance: 0.001t. 
$T_K$ follows universal scaling functions of
$1/(T^*L)$. Other parameters: $\epsilon_d$ = -0.8t, $t_R$ = $t_L$ = 0.4t, $\phi=0$. Here, we set $t=1$.}
\label{TK4n+2}
\end{figure}

To investigate the nature of the quantum phase transition 
in the thermodynamic limit more closely, we then perform the finite-size 
scaling for $T_k$ in the crossover region. 
We find $T_k$ in case (I) and (II) 
follows its own unique universal scaling function of $1/(T^*L)$ 
as $K\rightarrow K_c$ (see Fig. \ref{TK4n+2}), 
where $T^*$ is the crossover energy scale defined as:
 \begin{eqnarray}
T^*= c \widetilde{T_k}\exp{(-\pi\widetilde{T_k} /\mid K-K_c\mid)}
\label{Tstar}
\end{eqnarray} 
with $ \widetilde{T_k} =c' T_k$ has the same form in both case (I) and (II). 
Here, $c$, $c'$, $K_c$ are non-universal 
fitting prefactors depending on the initial parameters of the Hamiltonian. 
For $\epsilon_d$ = -0.8t, $t_R$ = $t_L$ = 0.4t, $\phi=0$, we find    
$c\approx 5.5$, $K_c\approx 0.0271$ (in unit of $t$) in both case (I) and (II);  
$c'\approx 0.65 $ in case (I), and $c'\approx 0.5$ for case (II). 
Note that the expression for $T^*$ in Eq. \ref{Tstar} 
is quite general as $K$ can be either smaller (case (I)) 
or larger (case (II)) than $K_c$. The universal scaling at finite sizes 
and the same exponential form for the crossover energy scale 
$T^*$ valid for both cases strongly indicate that in the thermodynamic limit 
the system exhibits the 
Kosterlitz-Thouless transition at a finite critical 
antiferromagnetic spin-spin coupling 
strength $K_c>0$. We have checked the consistency of our result 
from our finite-size scaling that $K_c$ indeed 
reaches to the same value in the thermodynamic limit 
for both case (I) and (II) even though the corresponding 
crossover regions are on the 
opposite side of the transition 
($K<K_c$ for case (I) and $K>K_c$ for case (II)). 

Note that unlike the similar side-coupled double-dot system studied 
previously in Ref.\cite{side-coupled} 
where the KT transition between  the Kondo and spin-singlet 
phase occurs at $K_c=0$, we find in our setup a finite $K_c>0$ 
for the same KT transition. 
We think this difference might be related to the more singular 
DOS of the conduction electron bath in the current setup of a tight-binding ring 
compared to that in a continuous Fermi sea with a constant DOS, making the Kondo 
resonance more robust against the antiferromagnetic spin-spin 
coupling in the current set up and therefore leads to 
a finite $K_c$ instead of $K_c=0$. However, it is also possible that 
the finite $K_c$ we find here 
is the artifact of the large-N slave-boson  mean-field theory.
 Further study is necessary to clarify this issue.

Finally, as a consistency check, as the system approaches to the thermodynamic limit, $L\rightarrow \infty$, $T_k\rightarrow T_k^0$ for $L=4n$ (case (I), 
$K<K_c$); while $T_k\rightarrow 0$ for $L=4n+2$ (case (II), $K>K_c$), as expected  (see Fig. \ref{TK4n+2}).

\subsubsection{{\bf C. Local Density of State}}
We now turn our attention to the local density of states (LDOS) 
on the dot $1$, given by
\begin{eqnarray}
&\rho_{QD}(\omega)& =-\frac{1}{\pi}Im{G^{R}}_{d_1d_1}(\omega)
\label{rhoQD}
\end{eqnarray} 
 where $G^{R}_{d_1d_1}(t)\equiv <d_1(t) d_1^{\dagger}(0)>$ 
is the retarded Green's function of dot $1$ which directly 
couples to the ring.
 Near the Fermi surface, $\rho_{QD}(\omega)$ determines 
the transport properties of the system, and it   
 can be obtained from the Green's function ${G}_{d_1d_1}(\omega)$ via 
equation of motion approach. 
First, the mean-field Hamiltonian in momentum space is given by:
\begin{eqnarray}
H_{MF} &=& \sum_{m,\sigma} \epsilon_m c_m^{\dagger}c_m +H.C. + \widetilde{\epsilon_d}f_{1\sigma}^{\dagger}f_{1\sigma}  - \sum_{m} (t_m c_m^{\dagger} 
f_1 + H.C.)\\\nonumber 
&-& \chi f_1^{\dagger}f_2 - H.C. + \frac{\chi^2}{K} + \lambda (b_0^2-1)
\end{eqnarray}
Where 
\begin{equation}
t_m = {\it i} \sqrt{\frac{2}{L}} \sin({\frac{m}{L}\pi}) [\widetilde{t_L} + \widetilde{t_R}\exp(i\phi)(-1)^{m+1}]
\end{equation}
 with $m=1,2,\cdots L-1$\cite{ring2}.

The equation for a general retarded Green's function $G_{ij}^R(\omega)$ is 
then given by:
\begin{eqnarray}
(\omega + {\it i} \eta-H_{MF})G_{ij}^R(\omega) = {\it I}
\end{eqnarray}
where ${\it I}$ is the identity matrix, 
$i,j$ can be $f_1$, $f_2$, or $m$. We therefore get the following three equations 
for $G_{f_1f_1}$, $G_{f_1m}$ and $G_{f_1f_2}$: 
\begin{eqnarray}
(\omega + {\it i} \eta-\widetilde{\epsilon_d})G_{f_1f_1}^R(\omega) - \chi G_{f_2f_1}^R - \sum_m t_m G^R _{md1}(\omega) =1
\end{eqnarray}
\begin{eqnarray}
- \chi G^R _{f_1f_1} + (\omega + {\it i} \eta)G^R _{f_2f_1} = 0
\end{eqnarray}
\begin{eqnarray}
- t_m G^R _{f_1f_1} + (\omega + {\it i} \eta - \epsilon_m) G^R _{mf_1} = 0
\end{eqnarray}
The above equations are easily solved and we get:
\begin{eqnarray} 
&{G}^R_{d_1d_1}(\omega)& = b_0^2 G^{R}_{f_1f_1} = 
\frac{b_0^2}{ \omega  -\widetilde{\epsilon_d} + {\it i} \eta
 \hspace{0.1cm}  - \sum_{m} [ \frac{t_{m}^2{\mid b_0 \mid}^2}{ \omega -\epsilon_m+{\it i} \eta}
] -  \frac{\chi^2}{ \omega +{\it i} \eta}}
\end{eqnarray}
where $\eta\rightarrow 0$. We then obtain $\rho_{QD}(\omega)$ by Eq. \ref{rhoQD}. 
In the following we analysize the crossover behaviors in 
LDOS $\rho_{QD}(\omega)$. 

\subsubsection{1. $K=0$}

\begin{figure}[h!]
\begin{center}
\includegraphics[width=14cm, height=12cm ]{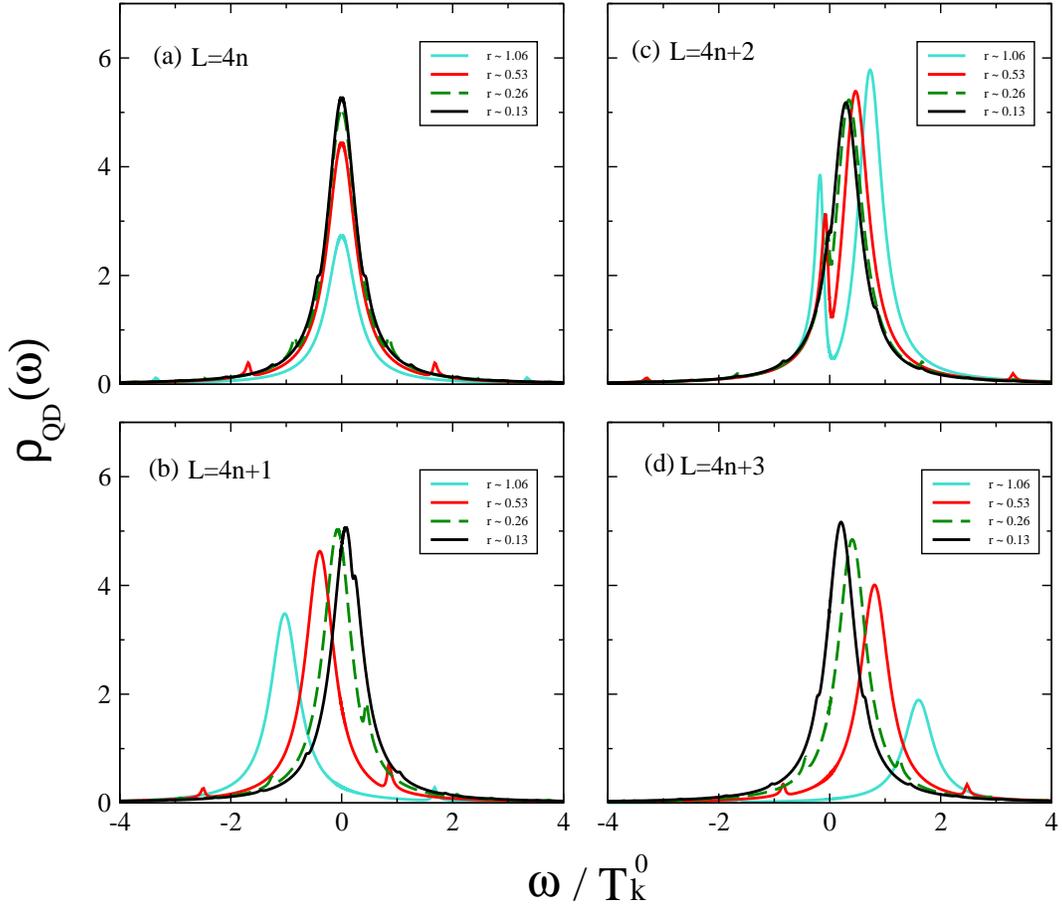}
\end{center}
\caption{ Local density of state for $K=0$ system (in arbitrary unit) (a) $L=4n$ 
(b) $L=4n+1$ (c) $L=4n+2$ (d) $L=4n+3$. Here, $r=\xi_k^0/L$, the ratio between 
$\xi_k^0$ and $L$. 
Other parameters: 
$\epsilon_d$ = -0.8t, $t_R$ = $t_L$ = 0.4t, $\phi=0$. Here, we set $t=1$.}
\label{0dos}
\end{figure}

In the absence of the antiferromagnetic spin-spin 
coupling ($K=0$) 
the LDOS has been extensively studied where 
$\rho_{QD}(\omega)$ depends sensitively 
on $L$ (mod$4$) at 
finite sizes \cite{ring2}.
As shown in Fig.(\ref{0dos}), our results on LDOS in this case 
 at finite sizes are qualitatively in good agreement 
with that in Ref.\cite{ring2}. 
For the convenience of later discussions we summarize below 
the behaviors of LDOS in the four different cases of $L$.
The key features are-- 
(i). There exsits a main Kondo resonance peak (located either 
symmetrically or asymmetrically 
with respect to $\omega=0$) followed by pairs of side peaks. 
(ii). As shown in Fig. \ref{0dos}, as $L$ increases 
the main Kondo peak gets more pronounced and closer to 
$\omega=0$; while the side peaks are gradually merged into the 
broadened main peak. 
In particular, 
the LDOS for $L=4n$ is very symmetric  
$\rho_{QD}(\omega)= \rho_{QD}(-\omega)$, suggesting the 
symmetry between particle and hole excitation energy 
in the finite size spectrum. The asymmetric Kondo peaks 
for $L=4n+1$ and $L=4n+3$ are located on the opposite 
sides (left for $L=4n+1$ and right for $L=4n+3$) 
of $\omega$.  
For $L=4n+2$, however, the LDOS shows 
asymmetric double Kondo peaks with comparable sizes and 
a dip at $\omega=0$. These  
differences in DOS among the four sizes of $L$ (mod $4$) 
can be explained in terms of the energy 
levels corresponding to the highest 
occupied (HO) and the lowest unoccupied (LU) states relative to the 
Fermi level\cite{ring2}: For $L=4n$, both HO and LU levels are 
around the Fermi energy $\epsilon_F=0$, leading to a symmetric single 
peak in LDOS at $\omega=0$. For $L=4n\pm 1$, HO and LU levels are 
both below and above $\epsilon_F$, respectively, giving rise 
to an asymmetric single 
peak in LDOS below and above $\omega=0$ separately. However, for $L=4n+2$, 
LO and HU levels are on the opposite side of the Fermi level, 
resulting in splitted double Kondo peaks below and above $\omega=0$.  
It should be noted that despite the differences at finite sizes 
($L\sim \xi_k^0$), the LDOS 
in all four cases indeed merges to a single Kondo peak as 
the system reaches the thermodynamic limit $L\rightarrow \infty$ 
(or $L\gg \xi_k^0$). 
(iii). The LDOS on dot $1$ obeys the following symmetries\cite{ring2}: 
$\rho_{QD}^{L}(\omega,\phi)= \rho_{QD}^{L+2}(\omega,\phi+\pi)$ 
and $\rho_{QD}(-\omega,\phi)= \rho_{QD}(\omega,\phi+\pi)$. 

We would like to point out here that from the width $D$ of the central 
Kondo peak(s), which can be approximately regarded as a quantity 
proportional to the effective Kondo 
temperature $T_k$, in LDOS 
for $K=0$ case one can qualitatively understand the opposite trends 
in $T_k$ at finite sizes in case (I) and (II) mentioned above. For 
$L=4n,4n+1,4n+3$ (case (I)), the width $D$ becomes larger as $L$ 
increases, indicating an increase in $T_k$ as the system 
approaches to the thermodynamic limit; while as for $L=4n+2$ (case (II)), 
$D$ gets smaller as $L$ increases, suggesting a decrease in $T_k$ 
as $L\rightarrow \infty$. 
We analysize in details below the behaviors of LDOS for $K>0$. 

\subsubsection{2. $K>0$}

At a finite $K>0$, 
the LDOS on the dot shows a crossover between 
the Kondo phase and the spin-singlet phase. 
For $K<K_{c1}$ at a fixed size $L$, the LDOS remains the same 
as that for $K=0$.  
For $L=4n,4n+2$ we find a continuous evolution in LDOS from 
Kondo to the crossover region near $K= K_{c1}$; while for 
$L=4n+1,4n+3$ the LDOS  
exhibits a first-order jump at $K_{c1}$. 
For $K>K_{c2}$, we find 
$\rho_{QD}(\omega) = 0$ as the indicator of the local spin-singlet 
phase since $b_0=0$. In the crossover region $K_{c1}<K<K_{c2}$, 
the Kondo peak in LDOS splits into two 
with respective to $\omega=0$. 
The splitting gets wider as $K$ increases 
further. Details are shown below.

\vspace{5mm}
\subsubsection{ {\large (i).} \textbf{ Varies $K$ with fixed $L$:}}
\vspace{5mm}

\begin{figure}[h!]
\begin{center}
\includegraphics[width=14cm, height=12cm ]{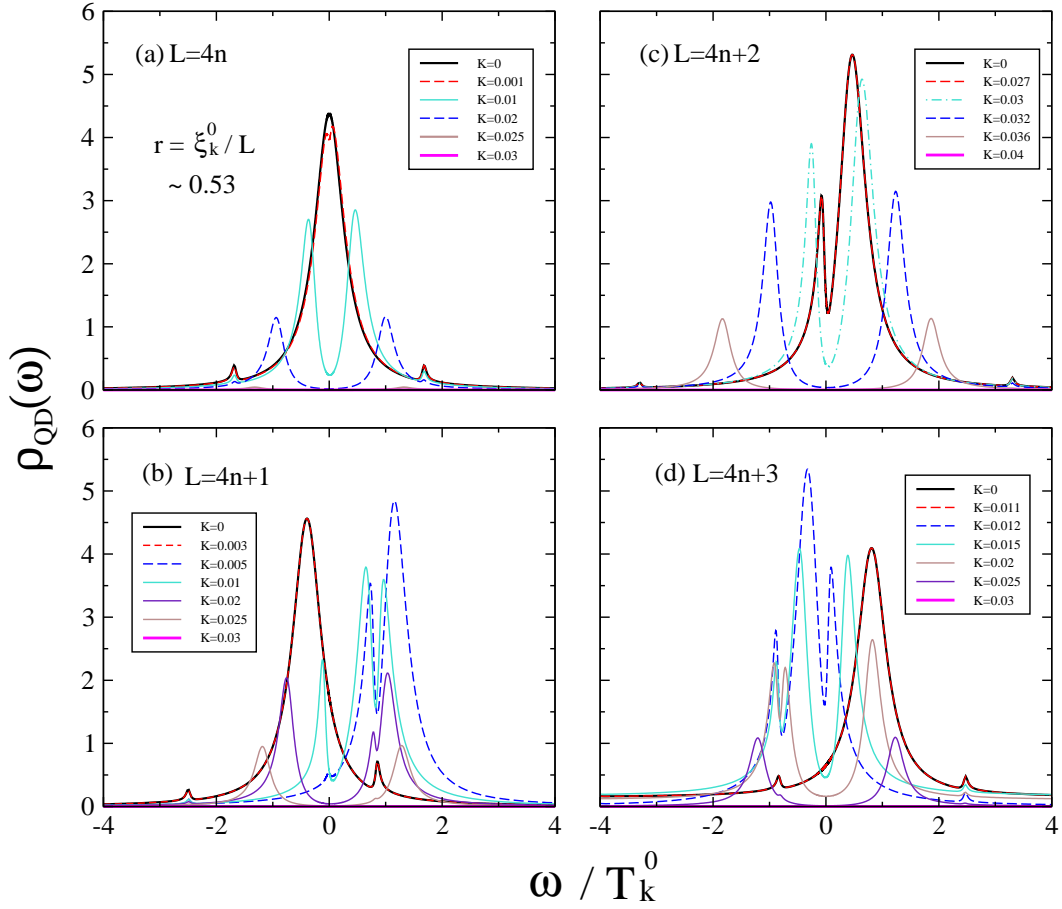}
\end{center}
\caption{Local density of state (in arbitrary unit) 
with various $K$ for $r=\xi_k^0/L\sim 0.53$. 
(a) $L=4n$ (b) $L=4n+1$ 
 (c) $L=4n+2$ (d) $L=4n+3$. Other parameters: $\epsilon_d$ = -0.8t, $t_R$ = $t_L$ = 0.4t, $\phi=0$. Here, $K$ is in unit of $t$, and we set $t=1$.}
\label{Kdos}
\end{figure}

In Fig.\ref{Kdos}, we show LDOS for various antiferromagnetic spin-spin 
coupling strength $K$ 
at a fixed size $L\approx 200$ ($\xi_k^0/L\sim 0.5$). 
The behaviors of LDOS are classified by $L$(mod $4$). 
The common features in Fig. \ref{Kdos} are as follows. 
In each of the four LDOS plots in Fig. \ref{Kdos}, the Kondo peak 
in $\rho_{QD}(\omega)$ splits into two 
at a small value of $K>K_{c1}$, indicating the crossover 
between Kondo and local spin singlet phases. The two peaks separate
further apart and become more symmetric as $K$ increases. At the
end, $\rho_{QD}(\omega)$ vanishes when $K>K_{c2}$ in the local spin 
singlet phase.
Note that the values for $K_{c2}$ depend sensitively 
on $L$ (mod $4$). This can be understood as when $L\approx 200$ 
the effective Kondo temperature $T_k$ in case (II) is much larger 
than that in case (I) (see Fig.\ref{TK4n}, Fig. \ref{TK4n+2}) , 
which explains why we need a larger value of $K_{c2}$ 
in case (II) to suppress the Kondo effect than that in case (I). 
 We present details below.

\vspace{5mm}
\subsubsection{ {\large (ii).} \textbf{ Varies $L$ with fixed $K$:}}
\vspace{5mm}

\begin{figure}[h!]
\begin{center}
\includegraphics[width=14cm, height=12cm ]{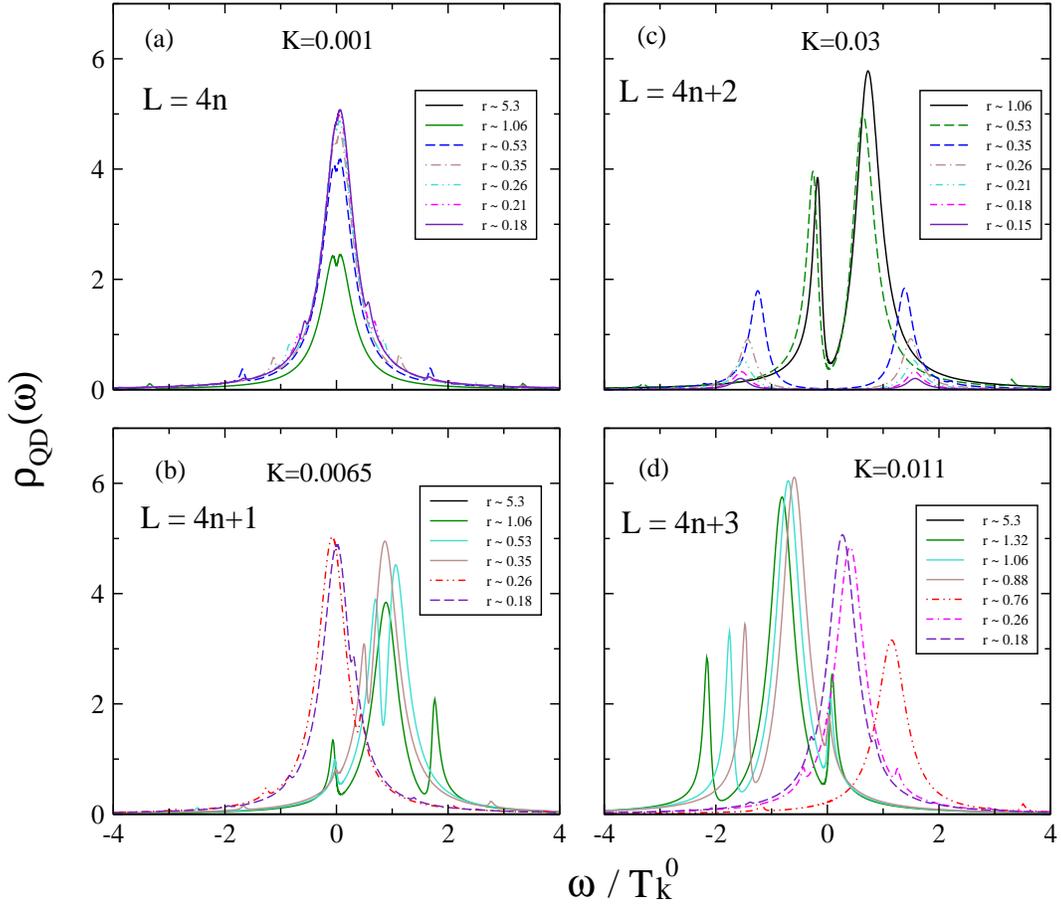}
\end{center}
\caption{Local density of state (in arbitrary unit) with various $L$ and fixed $K$ (in unit of $t$). Here, $r=\xi_k^0/L$. (a) $L=4n$ (b) $L=4n+1$ (c) $L=4n+2$ (d) $L=4n+3$. Other parameters: $\epsilon_d$ = -0.8t, $t_R$ = $t_L$ = 0.4t, $\phi=0$. Here, we set $t=1$.}
\label{dimdos}
\end{figure}

Fig. (\ref{dimdos}) shows how 
$\rho_{QD}(\omega)$ changes with the system size $L$. 
For $0<K<K_c$, starting from a small size $L<L_{c1}$, 
the LDOS changes from the behavior of a local spin-singlet 
state with $\rho_{QD}(\omega)=0$ to that in the crossover region 
with splitted peaks and finite LDOS at $\omega=0$,  
and finally it recovers the Kondo resonance at larger size $L>L_{c2}$.
 
We separate the discussion here into two different cases: 
$K<K_c$ (case (I)) and $K>K_c$ (case (II)). 
 Fig.(\ref{dimdos}(a),(b),(d))show finite size 
dependence of $\rho_{QD}(\omega)$ at $K<K_c$ (case (I)) for 
$L=4n$, $4n+1$, and $4n+3$, respectively. 
For small system sizes, the system is in the local spin-singlet state 
with vanishing LDOS. At intermediate sizes in the crossover region, however, 
LDOS starts to develop peaks with a dip at $\omega=0$. As size further 
increases, these splitted peaks either gradually ($L=4n$) or 
suddenly ($L=4n+1$ and $4n+3$) merge into a single Kondo peak 
located either symmetrically ($L=4n$) or asymmetrically ($L=4n+1$ and $4n+3$) 
with respect to $\omega=0$. These Kondo peaks then follow 
the evolution at finite sizes for $K=0$ and finally recover the 
single symmetric Kondo peak in the thermodynamic limit.  

For $K>K_c$ (case (II)), the size dependence of
$\rho_{QD}(\omega)$ is shown in
Fig.(\ref{dimdos}(c)) with $K=0.03t > K_c$. 
The LDOS exhibits a crossover from the Kondo to local spin-singlet 
state with increasing size $L$. 
For $K \sim K_c$ the LDOS changes from
singlet state at small sizes to the crossover regime within 
the range $100 <L< 800$ that we investigate. To study the 
the ultimate fate of the system one needs to go much larger 
system size than $L\sim 800$, which is beyond the scope of 
our computational limit. 
 Note that our analysis on the LDOS by changing 
$K$ with fixed $L$ (changing $L$ with fixed $K$) is consistent 
 with the green line (red line) of the 
schematic phase diagram shown in Fig. (\ref{phase}). 
Finally, the first order jumps seen in LDOS at $K=K_{c1}$ 
for $L= 4n+1$ and $L=4n+3$ may be due to the artifact of the 
mean-field theory. Further investigation is needed to clarify 
this issue. 

\subsubsection{{\bf D. Persistent Current}}

We now analysize the crossover from the behaviors in persistent current. 
The Kondo screening cloud in our closed setup is restricted itself 
in the ring. To get an experimental access, one possible way is to measure 
persistent current (PC) induced by changing the magnetic flux threading
the ring without attaching leads to it. PC is defined as: 
\begin{equation}
I=-\frac{e}{\hbar} \frac{\partial
E_{gs}}{\partial \phi}
\end{equation}
PC can be served as a detector for the Kondo screening cloud since 
as a function of $\phi$ PC behaves very differently when $\xi_k \gg L$ 
than when $\xi_k\ll L$. 
Such persistent current experiments have been reported recently on micron sized ring without containing the dot\cite{PCexp}. In our setup, PC can be used 
to measure the strength of the Kondo correlation as the antiferromagnetic 
coupling $K$ is expected to suppress the Kondo effect, leading to a 
weaker PC.

\subsection{K=0}

\vspace{5mm}
\begin{figure}[h!]
\begin{center}
\includegraphics[width=13cm]{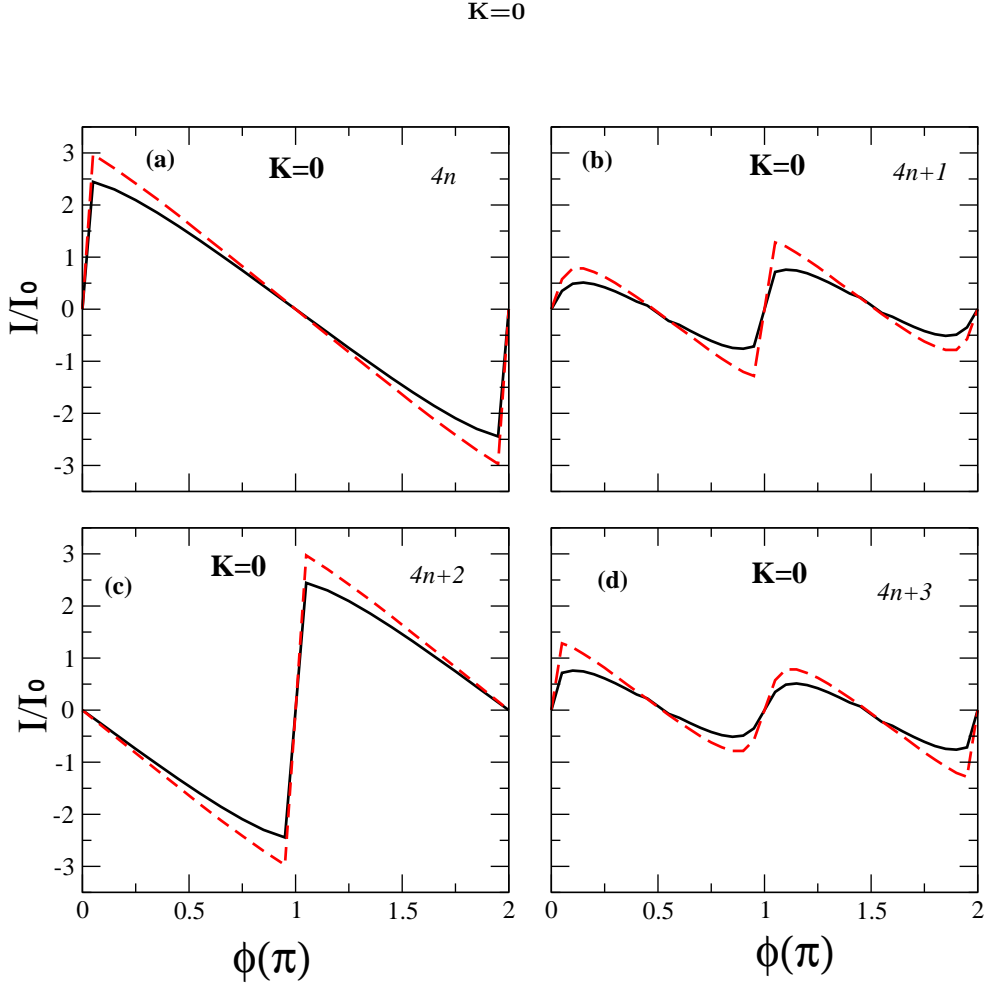}
\end{center}
\caption{Persistent current (in unit of $I_0$) 
versus magnetic flux (in unit of 
$2\pi/\Phi_0$) with $K = 0$. 
Red dashed line is for $r=\xi_k^0\sim 0.53$. Black solid line is for 
$\xi_k^0/L\sim 1.06$. (a) $L=4n$ (b) $L=4n+1$ (c) $L=4n+2$ (d) $L=4n+3$. Other parameters: $\epsilon_d$ = -0.8t, $t_R$ = $t_L$ = 0.4t. Here, we set $t=1$. }
\label{pc0}
\end{figure}

Before we present our new results on the PC, it proves 
useful to summarize the previous 
results\cite{ring1}\cite{ring2} for $K=0$. 
As shown in Fig.(\ref{pc0}), since the system is in the Kondo regime, 
PC increases with increasing system size $L$. 
PC as a function of the magnetic flux is a weak sinusoidal wave 
when $L \ll \xi_k^0$; while it 
behaves like a saw-tooth  
when $L \gg \xi_k^0$. Meanwhile, there exists a relation between 
PC for $L$ and $L+2$: $I_N(\phi) = I_{N+2}(\phi + \pi)$\cite{ring1} 
\cite{ring2} due to the $\pi$ shift in LDOS of the quantum dot from 
the system with $N$ sites to $N+2$ sites. Furthermore, 
the magnitude of PC for $L=odd$ is much smaller than that for $L=even$. 
This can be understood as for $L=odd$ the electrons are fully occupied 
at the Fermi level, which suppresses the Kondo resonance; while as for $L=even$ 
there is an unpaired electron on the Fermi level, leading to the 
Kondo resonance. The results for $K>0$ are shown below.

\vspace{5mm}
\subsubsection{\textbf{1. Varies $K$ with fixed $L$:}}
\vspace{5mm}

\begin{figure}[h!]
\begin{center}
\includegraphics[width=14cm, height=12cm]{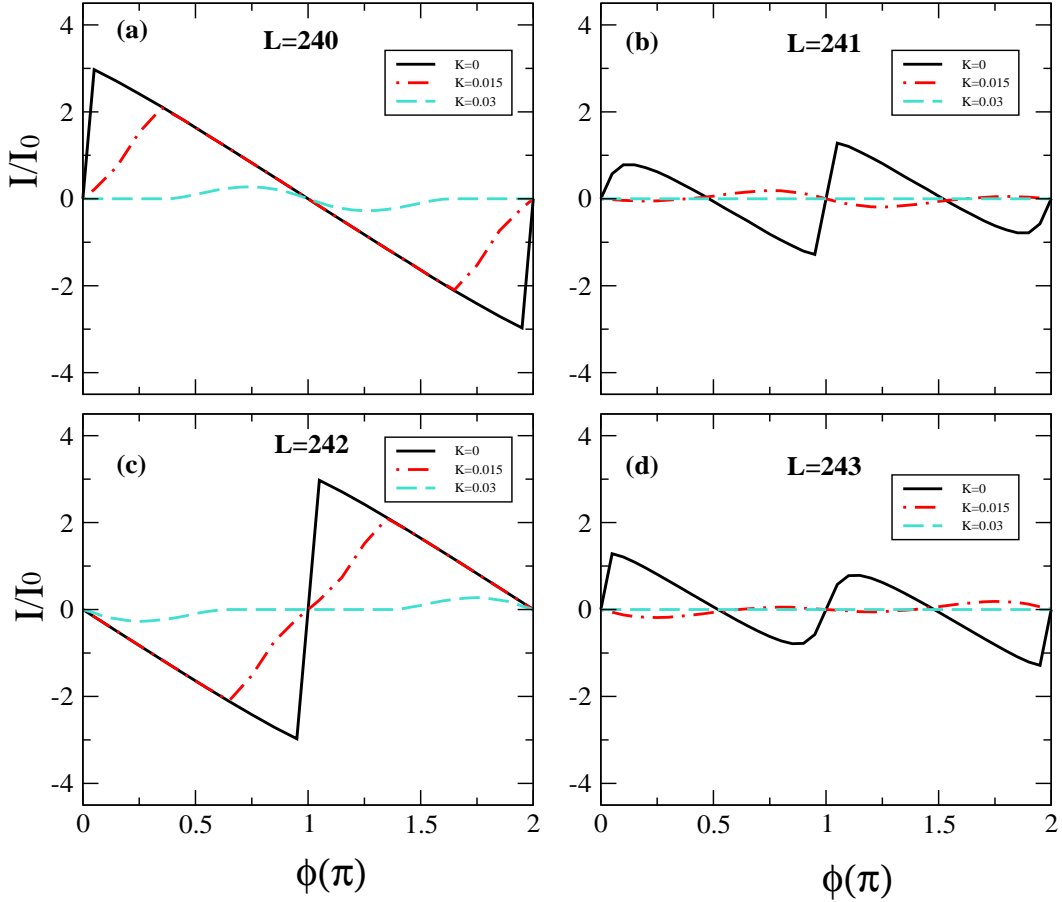}
\end{center}
\caption{PC versus $\phi$ with various $K$ for $L\sim 240$ 
($\xi_k^0/L\sim 0.44$). $K = 0, 0.015, 0.03$ 
(in unit of $t$) (a) $L=4n$ (b) $L=4n+1$ (c) $L=4n+2$ (d) $L=4n+3$. Other parameters of data: $\epsilon_d$ = -0.8t, $t_R$ = $t_L$ = 0.4t. Here, we set $t=1$.}
\label{kpc}
\end{figure}

 PC (in unit of $I_0$ 
where $I_0=e v_F/L$ is the persistent current of an ideal metallic ring 
with $v_F$ being Fermi velocity) versus magnetic flux with different antiferromagnetic 
coupling strength $K$ at a finite size 
$L\sim 240$ ($\xi_k^0/L\sim 0.44$) is shown in Fig.(\ref{kpc}). 
The general trend in all four cases is that the 
amplitude of PC gets smaller as $K$ is increased and finally 
vanishes for $K>K_{c2}$, in agreement 
with the expectation on the suppression of the Kondo effect 
by the antiferromagnetic spin-spin 
coupling. Also, as shown in Fig. \ref{kpc}, 
the shape of the PC as a function of $\phi$ changes 
from saw-tooth shape at smaller $K$ values 
to weak sinusoidal waves at 
larger $K$ values, similar to that for $K=0$. 
There are detail differences among 
the four cases. 
In Fig.(\ref{kpc}(a) and (c)), with increasing $K$ the persistent 
current $I_ {L=240}$ near $0.5\pi< \phi<1.5 \pi$
decreases more slowly than that for $0\pi <\phi <0.5\pi$ 
and $1.5\pi <\phi <2\pi$; similarily for $I_{L=242}$ except for the range 
of $\phi$ is being interchanged. 
This suggests that the Kondo effect is more robust in 
$0.5\pi< \phi<1.5 \pi$ for $L= 4n$ 
and in $0\pi <\phi <0.5\pi$ 
and $1.5\pi <\phi <2\pi$ for $L=4n+2$. 
Note that we find the similar  
jumps at $K=K_{c1}$ mentioned previously in LDOS to also 
appear in PC for $L=4n+1$ and $L=4n+3$; while PC is continuous 
at at $K=K_{c1}$ for the other two cases. 
Despite the above detail differences, we 
find two common features in Fig. (\ref{kpc}) 
that remain the same as in the case of $K=0$: 
 (i). $I_N(\phi) = I_{N+2}(\phi + \pi)$ and (ii). 
PC for $L=even$ is larger than that for $L=odd$. 
\vspace{5mm}

\subsubsection{\textbf{2. Varies $L$ with fixed $K$:}}
\vspace{5mm}

\begin{figure}[h!]
\begin{center}
\includegraphics[width=14cm, height=12cm ]{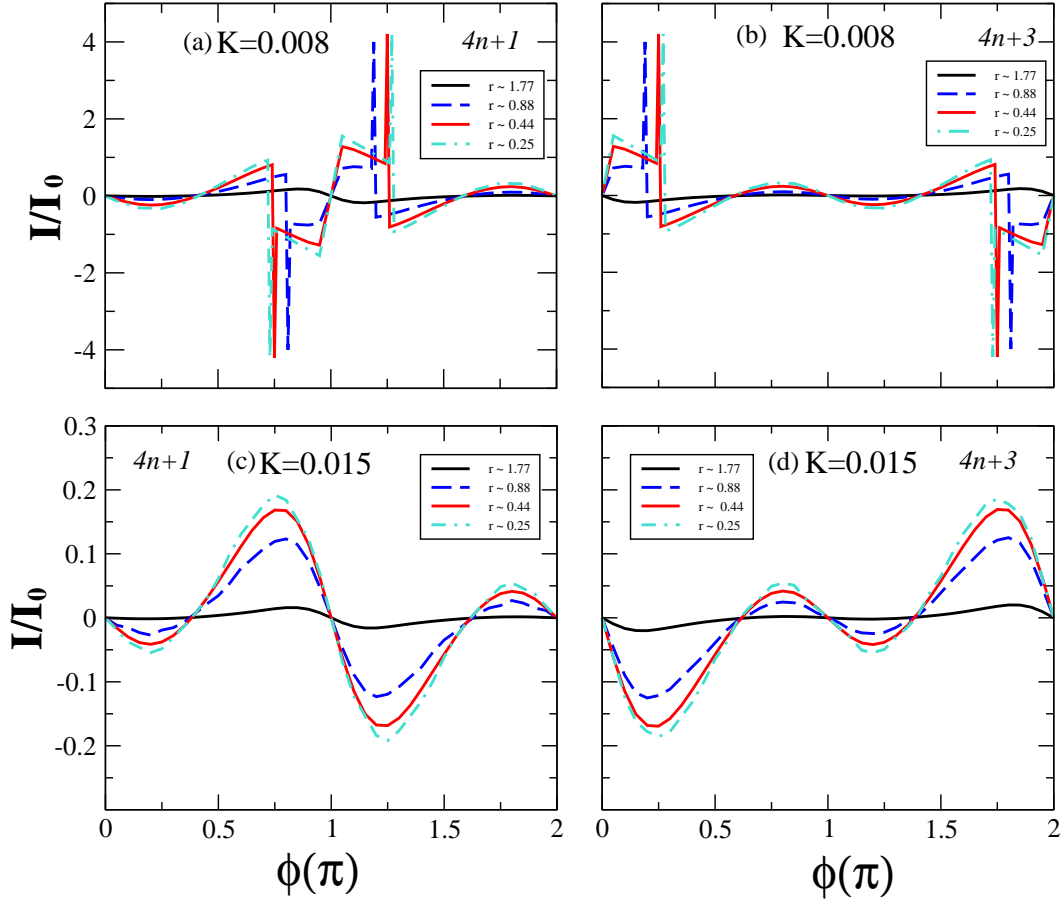}
\end{center}
\caption{Persistent current with various $L$ and fixed $K$ (in unit of $t$). 
(a) $L=4n+1$, $K=0.008$ (b) $L=4n+3$, $K=0.008$ (c) $L=4n+1$, $K=0.015$ (d) $L=4n+3$, $K=0.015$. Here, $r=\xi_k^0/L$. The parameters are: $\epsilon_d$ = -0.8t, $t_R$ = $t_L$ = 0.4t, $\phi=0$. Here, we set $t=1$.}
\label{pcodd}
\end{figure}

\begin{figure}[h!]
\begin{center}
\includegraphics[width=14cm, height=6cm ]{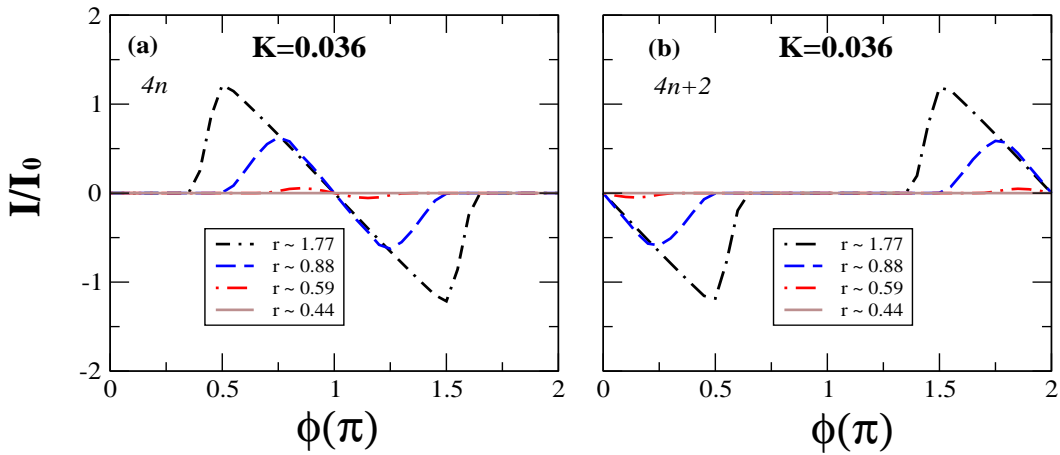}
\end{center}
\caption{Persistent current with various $L$ and fixed $K$. 
(a) $L=4n$ (b) $L=4n+2$. Here, $r=\xi_k^0/L$. Other parameters of data: $\epsilon_d$ = -0.8t, $t_R$ = $t_L = 0.4t$, $K=0.036$ (in unit of $t$). Here, we set $t=1$.}
\label{pceven}
\end{figure}

Fig.(\ref{pcodd}) and Fig.(\ref{pceven}) show PC versus 
magnetic flux at different system sizes. 
 In the following the discussion is separated into two cases: 
(i). $K < K_c$ (case (I)) where PC increases with increasing size $L$, and 
(ii). $K > K_c$ (case (II)) where PC decreases to $0$ as $L$ reaches the thermodynamic limit.

\begin{center}
\bf{(i). $K \le K_c$}
\end{center}

\vspace{5mm}
We first look at cases with $K<K_c$. 
Fig.(\ref{pcodd}) shows PC versus magnetic flux with different size $L$ 
for $K=0.008$ and $K=0.015$ (in unit of $t$). 
For $K=0.008$ (see Fig.\ref{pcodd}(a)(b)), 
as $L$ is increased we find the PC exhibits two different 
behaviors: near $\phi=\pi$ for $L=4n+1$ 
and  $\phi=0,2\pi$ for $L=4n+3$ PC changes from 
the behavior in crossover region to the Kondo state at $K=0$ 
(see Fig. (\ref{pc0}(b)(d)); while out of these ranges of $\phi$  
its behavior remains the same as in the crossover region but 
with an increasing amplitude. A first-order jump 
is seen to separate these two regions. We expect at much larger system 
size the Kondo phase  
will be restored eventually over the entire range of $\phi$.   
We have also investigated PC for $L=4n,4n+2$ (not shown here) 
and find the same qualitative behaviors except that instead of 
jumps we find a continuous change in PC. 
Also, the behavior in 
$I_{4n+1}^{ \phi \sim \pi}$ (or equivalently $I_{4n+3}^{  \phi \sim 0}$) 
is the same as that for $K=0$ for all the sizes we 
investigate, suggesting the Kondo phase is very easily 
restored in these ranges of $\phi$.

On the other hand, for a larger value of 
$K=0.015$ (see Fig.\ref{pcodd}(c)(d)), as $L$ increases from $L\sim 60$ to $L\sim 420$ (or $r=\xi_k^0/L$ changes from $1.77$ to $0.25$) 
PC still stays in the crossover region with 
 an increasing amplitude. This is expected as for larger value of $K$ 
one must go to much larger system size to 
observe the restoring of the Kondo effect in PC.

\begin{center}
\bf{(ii). $K \ge K_c$}
\end{center}

\vspace{5mm}

We now investigate the case where $K>K_c$. 
As shown in Fig.(\ref{pceven}), since $K>K_{c}$ in this case, 
PC decreases in amplitude with increasing 
system size $L$ and finally the system reaches the 
local spin-singlet state with vanishing PC. 
Note that $I_{4n}^{\phi\sim 0}$ and correspondingly 
$I_{4n+2}^{\phi\sim \pi}$ 
are always vanishingly small for all the sizes we investigate, 
suggesting the systems are already 
in the local spin-singlet states from the start 
for these ranges of  $\phi$, which is consistent with 
our mean-field phase diagram. However, 
for the remaining ranges of $\phi$ 
the systems start from the crossover region for smaller sizes and 
reach finally to the local spin-singlet state at large size. 

From above results for PC, it implies that by applying magnetic flux, 
we may change the ground state 
of our system at finite sizes either from the local spin-singlet 
state to the crossover region or from the Kondo phase to the crossover 
region.

\subsection{IV. Conclusions}
To summarize, we have studied via large-N slave-boson mean-field approach 
the Kondo effect in a side-coupled double-quantum-dot system where 
one dot is embedded in a mesoscopic ring. The competition between the 
Kondo effect and the antiferromagnetic spin-spin 
interaction in this geometry gives rise to 
the Kosterlitz-Thouless quantum phase transition with a finite critical 
value $K_c$ in the thermodynamic limit. 
The mean-field phase diagrams of the model depends on the finite size $L$ (mod $4$): 
for $L=4n,4n+1,4n+3$ (case (I)) 
the crossover occurs for $K<K_c$; while for $L=4n+2$ (case (II)) the 
crossover exists for $K>K_c$. To further study how the Kondo screening is suppressed 
by the RKKY, we have performed a systematic finite-size analysis on the Kondo temperature $T_k$, the 
local density of states on the dot $\rho_{QD}(\omega)$ which connects to the ring , and 
the persistent current PC induced by the magnetic flux penetrating through the ring.
For a fixed $K<K_c$, we find all the above quantities flow to the Kondo phase (same phase as in $K=0$) 
with increasing 
system size: 
$T_k$ and PC increase in magnitude and $\rho_{QD}(\omega)$ develops a pronounced 
single Kondo peak centered at $\omega=0$; while for $K>K_c$, the flow at finite sizes 
is towards to the local 
spin-singlet phase where all of the three observables decrease in magnitude 
to $0$ with increasing size. 
From the finite-size scaling of $T_k$, we have shown that $T_k$ is an universal function 
of $1/(LT^*)$ for $1/L<T^*$ where $T^*$ has been unambiguously identified the characteristic 
 Kosterlitz-Thouless crossover energy scale:  
$T^*= c \widetilde{T_{k}}\exp{[-\pi\widetilde{T_{k}}/(K-K_c)]}$. 
For a fixed size $L$, we find the ground state remains in the Kondo phase for $K<K_{c1}$ and 
it crossovers to the local spin-singlet phase for $K_{c1}<K<K_{c2}$ and finally reaches the 
spin-singlet phase for $K>K_{c2}$. For $L=4n+1,4n+3$, we find the first order jumps in 
all of the three observables at the phase boundary between Kondo and the crossover region. 
Whether these first order jumps are the artifacts of the large-N mean-field theory is yet 
to be clarified. Note that unlike the similar system studied previously on the 
side-coupled double-dot system embedded in conduction electron Fermi sea with constant density 
of states, 
the key finding in this paper is the KT transition with 
a finite critical value $K_c>0$ in a side-coupled 
double-dot system embedded in a mesoscopic ring (in stead of $K_c=0$ for the 
system with Fermi sea of continuous spectrum). 
Whether this finite $K_c$ is due to the artifact of the mean-field theory or due to the more  
singular density of states of the 1D tight-binding ring needs further investigations. Our results on the transport properties of the system 
are relevant for future experiments on the side-coupled 
double-quantum-dot system embedded in a mesoscopic ring.\\

\acknowledgements

We are greatful for the useful discussions with 
G.M. Zhang, C.S. Chu, J.J. Lin 
and J.C. Chen.  We also 
acknowledge the generous support from the NSC grant
No.95-2112-M-009-049-MY3, the MOE-ATU program, the NCTS of Taiwan, R.O.C., 
and National Center for Theoretical Sciences (NCTS) of Taiwan.

\end{document}